\documentclass[12pt]{article}
\usepackage{amssymb,amsmath,epsfig}
\usepackage{graphicx}
\usepackage{subfig}
\usepackage{cite}
 \allowdisplaybreaks
\begin{document}
\title{\bf Energy Conditions in Non-local Gravity }

\author{M. Ilyas\thanks{ilyas\_mia@yahoo.com}\\
Centre for High Energy Physics, University of the Punjab,\\
Quaid-i-Azam Campus, Lahore-54590, Pakistan}

\date{}

\maketitle
\begin{abstract}
We investigate the different energy conditions in nonlocal gravity, which is obtained by adding an arbitrary function of d'Alembertian operator, $f(\Box^{-1})$, to the Hilbert- Einstein action. We analyze the validity of four different energy conditions and illustrate the different constraints over parameters of the power law solution as well as de Sitter solution.
\end{abstract}

\section{Introduction}

As a matter of facts, a couple of models survive to explain the latetime cosmic acceleration of universe. One of them is $\Lambda$CDM \cite{ref5}, which is simple, popular and agree more with the cosmological data. The role of dark energy (DE) is due to the cosmological constant ($\Lambda$) which is added to Hilbert- Einstein action. Some other models assume the dynamical DE described by varying parameter of EoS. In order to get a varying parameter of EoS, the standard procedure is to introduce the scalar fields to cosmological models. In fact, the cosmological models with scalar fields can well describe the mechanism behind the evolution of universe, e.g. the Quintom model having two scalar fields: an ordinary scalar field as well as phantom scalar field \cite{ref9}. Quintom cosmological models are an active field of research, for further detail, see \cite{ref10,ref11,ref12,ref13,ref14,ref14a,ref15}.\\
Apart from this, another suitable approach is the modified theories of gravity.
The modified gravitational theories are more widely applicable models which are held by modifying the gravitational part of Hilbert-Einstein action.  Nojiri and Odintsov studied the behavior of cosmic accelerating nature of universe under modified $f(R)$ gravity. Some of modified gravity and dark matter are studied in ref \cite{refa1,refa2,refa3,refa4,refa5,refa6,refa7,refa8,refa9,ref9a,refa10,refa11,refa12}.
\\
In the context of quantum gravity, some higher derivative correction terms are added to Hilbert- Einstein action which are studied in ref \cite{ref23,ref24}. Due to quantum effects, the non-local gravity theories were suggested \cite{ref25}. According to string/M theory, which is widely supposed as an essential theory of all known interactions including gravity, as a result, a natural appearance of non-locality in the background of string field theory offers a completely motivational study for non-local cosmological models. Most of the non-local gravity theories contain a non-local scalar field coupled with gravity or a function of d'Alembered operator ($\Box$) or, in Hilbert- Einstein action and considered as a modified gravity \cite{ref26,ref26a,ref26b,ref26c,ref27,ref28,ref29,ref30,ref31,ref32,ref32a,ref32b,ref33,ref34,ref35,ref36,ref37,ref38,ref39,ref40,ref41,ref41a}.
\\
The energy conditions are the fundamental elements as the esoteric study of the singularity theorem. Hawking-Penrose singularity theorem brought down the importance of the weak and strong energy conditions, although the black hole second law of thermodynamics indicate null energy condition. The familiar Raychaudhuri equations gave notice of the viability of several variants of energy conditions. The generalized energy conditions in modified theories of gravity were studied in ref\cite{ref42,ref42a}, in which the authors discussed the formulation and the meaning of the energy conditions in the context of modified theories of gravity.
\\
In this paper, we will focus our investigation on non-local gravity models. We then apply some certain constraints over non-local gravity models with the help of energy conditions and showed that these models can satisfy energy conditions in specific region under power-law solutions. This paper is organized that the next coming section consists of a brief introduction to non-local gravity models and its equation of motion along with the effective version of all energy conditions. In section 3, we demonstrate the energy conditions through different plots for two type of solutions and the last section consists of summarized results and conclusions.
\section{A class of non-local gravity}
\label{origin} We start from a class of non-local
gravities, with the following action
\begin{equation}
\label{action} S=\int d^4 x \sqrt{-g}\left\{
\frac{1}{2\kappa^2}\left[ \mathcal{R}\left(1 + f(\Box^{-1}\mathcal{R})\right) -2\lambda
\right] + \mathcal{L}_\mathrm{m} \right\}\, ,
\end{equation}
where $\kappa^2=8\pi/{m_{\mathrm{Pl}}}^2$, the Planck mass read as $m_{\mathrm{Pl}} =  1.2 \times 10^{19}$ GeV, while $f$,  $\Box^{-1}$, $\lambda$ and $\mathcal{L}_\mathrm{m}$ are  differentiable  function,  the inverse of the d'Alembertian operator, cosmological constant, and the matter Lagrangian, respectively. The covariant d'Alembertian for a scalar field, $\eta$, which reads
\begin{equation*}
\Box \eta\equiv \nabla_\mu\nabla^\mu\eta= \frac{1}{\sqrt{-g}} \partial_\mu
\left[ \sqrt{-g} \, g^{\mu
  \nu}\partial_\nu \eta\right],
\end{equation*}
where $\nabla_\mu$ is the covariant derivative and $g$ is the determinant of the metric tensor,
$g_{\mu\nu}$.

For the localization procedure, we introducing two scalar fields, $\phi$ and $\xi$. We define
$\phi=\Box^{-1}\mathcal{R}$ and $\xi$ as Lagrange multiplier. Using these definitions in action~(\ref{action}), we get a
local action, written as
\begin{equation}
\label{action2} S_L= \int d^4x \sqrt{-g}\left[
\frac{1}{2\kappa^2}\left[ \mathcal{R}\left(1 + f(\phi)\right) +\xi
\left(\mathcal{R}-\Box \phi\right) - 2 \lambda \right]+ \mathcal{L}_\mathrm{m}
\right] \, .
\end{equation}
Now to get the field equation for scalar fields, one need to vary this action with
respect to $\xi$ and $\phi$, as a result
\begin{eqnarray} \label{equpsi} \Box\phi&=&\mathcal{R}\,,
\\
\Box\xi&=& f_{,\phi}(\phi) \mathcal{R}\, , \label{equxi} \end{eqnarray} where
$f_{,\phi}(\phi)=\frac{df}{{d}\phi}$. Furthermore, the Einstein field equations can be obtained by varying the
action~(\ref{action2}) with respect to $g_{\mu\nu}$, as follows
\begin{equation}
\label{nl4}
\begin{split}
&\frac{1}{2}g_{\alpha\beta} \left[\mathcal{R}\Psi
 + \partial_\rho \xi \partial^\rho \phi - 2 (\lambda+\Box\Psi) \right]
 - R_{\alpha\beta}\Psi-\frac{1}{2}\left(\partial_\alpha \xi \partial_\beta \phi
+ \partial_\alpha \phi \partial_\beta \xi \right)
  + \nabla_\alpha \partial_\beta\Psi=- \kappa^2 T^{(m)}_{\alpha\beta},
\end{split}
\end{equation}
where $\Psi\equiv1+f(\phi) +\xi$, and $T^{(m)}_{\alpha\beta}$ is
the energy--momentum tensor of the matter sector, defined as
\begin{equation}\label{Tmatter}
T^{(m)}_{\alpha\beta} \equiv -\frac{2}{\sqrt{-g}}
 \frac{\delta \left(\sqrt{-g} \mathcal{L}_\mathrm{m}\right)}{\delta g^{\alpha\beta}}.
\end{equation}

We can rewrite the above equation as $G_{\mu \nu}=T_{\mu \nu}^{eff}$ where
$$G_{\mu \nu}={R_{\mu \nu }} - \frac{1}{2}{g_{\mu \nu }}\mathcal{R}$$
and
\begin{align}
&T_{\mu \nu }^{eff} =\kappa^2 T_{\mu \nu}  - \frac{1}{2}({\partial _\mu }\xi {\partial _\nu }\left[ {1 + f(\phi ) + \xi } \right] + {\partial _\mu }\left[ {1 + f(\phi ) + \xi } \right]{\partial _\nu }\xi )\nonumber\\
& + {\nabla _\mu }{\partial _\nu }\left[ {1 + f(\phi ) + \xi } \right] - {g_{\mu \nu }}(\Lambda  +\Box \left[ {1 + f(\phi ) + \xi } \right]\nonumber\\
& + \frac{1}{2}{g_{\mu \nu }}{\partial _\rho }\xi {\partial ^\rho }\left[ {1 + f(\phi ) + \xi } \right] - \frac{1}{2}{g_{\mu \nu }}\mathcal{R}\left[ {f(\phi ) + \xi } \right] - \left[ {f(\phi ) + \xi } \right]{R_{\mu \nu }}.
\end{align}
In this paper, we assume FLRW universe model, with the line element
\begin{eqnarray}\label{FLRW}
ds^2={}-dt^2+a^2(t)\delta_{ij}dx^idx^j\,.
\end{eqnarray}
We consider that the perfect fluid and assume the simple case where both $\phi$ and $\xi$
are the time dependent functions. Thus, the system of
Eqs.~(\ref{equpsi})--(\ref{nl4}) reduces to

\begin{equation}
\begin{split}
&{\rho ^{eff}} = {\kappa ^2}{\rho _{\rm{m}}} - \frac{1}{2} \xi' \phi'  - 3H \Psi'  + \lambda {\mkern 1mu}  - 3{H^2}(f(\phi ) + \xi ),\\
&{P^{eff}} = {\kappa ^2}{P_{\rm{m}}} - \frac{1}{2} \xi' \phi'  +  \Psi''  + 2H \Psi'  - \lambda {\mkern 1mu}  + \left( {2 H' + 3{H^2}} \right)(f(\phi ) + \xi ),\\
& \phi''  =  - 3H \phi'  - 6\left( { H' + 2{H^2}} \right){\mkern 1mu} ,\\
& \xi''  =  - 3H \xi'  - 6\left( { H' + 2{H^2}} \right){f_{,\phi }}(\phi ){\mkern 1mu} ,
\end{split}
\end{equation}

where prime means differentiation
with respect to time, $t$, and $H= a'/a$ is the Hubble
parameter.

\subsection{Energy conditions}
Energy conditions are the mathematical constraint which can be applied to energy momentum tensor (matter content) for the purpose to check either the matter is physically acceptable or not. These are the coordinates invariant and play an important role in the study of black hole e.g. the laws of blackhole thermodynamics and no hair theorem. The idea came from Raychaudhuri equations, as follow
\begin{eqnarray}\label{10}
\frac{d\theta}{d\tau}&=&\Omega^{\alpha\beta}
\Omega_{\alpha\beta}-\Sigma^{\alpha\beta}\Sigma_{\alpha\beta}-\frac{1}{3}\theta^2
-R_{\alpha\beta}u^{\alpha}u^{\beta},\\\label{11}\frac{d\theta}{d\tau}
&=&\Omega^{\alpha\beta}\Omega_{\alpha\beta}
-\Sigma^{\alpha\beta}\Sigma_{\alpha\beta}-\frac{1}{2}\theta^2-R_{\alpha\beta}k^{\alpha}
k^{\beta},
\end{eqnarray}
The above equations are the temporal variations of expansion scalar, $\theta$, in which $\Omega_{\alpha\beta},~\Sigma_{\alpha\beta},~u^{\alpha}$ and
$k^{\alpha}$ represent the rotation, shear tensor, timelike and null
tangent vectors, respectively. \\
By neglecting the rotation and distortions due to quadratic terms and simplifying Eqs.(\ref{10}) and
(\ref{11}), we reach
\begin{equation}\nonumber \theta=-\tau
R_{\alpha\beta}u^{\alpha}u^{\beta},\quad \theta=-\tau
R_{\alpha\beta}k^{\alpha}k^{\beta}.
\end{equation}
As gravity is attractive by nature which means $\theta<0$, so we get $R_{\alpha\beta}u^{\alpha}u^{\beta}\geq0$ and
$R_{\alpha\beta}k^{\alpha}k^{\beta}\geq0$. Similarly, with the help of Einstein
field equations, the other expression for these inequalities can be written  as
\begin{equation}\nonumber
\left[T_{\alpha\beta}-\frac{1}{2}g_{\alpha\beta}T\right]u^{\alpha}u^{\beta}\geq0,\quad
\left[T_{\alpha\beta}-\frac{1}{2}g_{\alpha\beta}T\right]k^{\alpha}k^{\beta}\geq0.
\end{equation}
Here we deal alternate to Einstein theory of gravity so the energy conditions can be formulated by replacing the ordinary energy momentum tensor, $T_{\alpha,\beta}$, by effective energy momentum tensor, $T^{eff}_{\alpha,\beta}$.\\
For perfect fluid matter distribution, these inequalities provide
the energy constraints defined by
\begin{itemize}
\item NEC:$\quad\rho^{eff}+p^{eff}\geq0$,
\item WEC:$\quad\rho^{eff}+p^{eff}\geq0,\quad\rho^{eff}\geq0,$
\item SEC:$\quad\rho^{eff}+p^{eff}\geq0,\quad\rho^{eff}+3p^{eff}\geq0,$
\item DEC:$\quad\rho^{eff}\pm P^{eff}\geq0,\quad\rho^{eff}\geq0.$
\end{itemize}
We can see that the NEC is fundamental in all other and its violations leads the violation of all other conditions.\\
These conditions take the following form
\begin{itemize}
\item \textbf{NEC}:$\quad\rho^{eff}+p^{eff}=p + \rho  + 2f(\phi )H' + 2\xi H' - \xi '\phi ' - H\Psi ' + {\Psi ^{\prime \prime }}\geq0$,
\item \textbf{WEC}:$\quad\rho^{eff}=\rho  - 3f(\phi ){H^2} - 3{H^2}\xi  - \frac{1}{2}\xi '\phi ' - 3H\Psi ' + \lambda\geq0,$
\item \textbf{SEC}:$\quad\rho^{eff}+3p^{eff}=\rho  + 3p - 2\lambda  + 6f(\phi ){H^2} + 6{H^2}\xi  + 6f(\psi )H' + 6\xi H' - 2\xi '\phi '
    + 3H\Psi ' + 3{\Psi ^{\prime \prime }}\geq0,$
\item \textbf{DEC}:$\quad\rho^{eff}- p^{eff}=\rho  - p + 2\lambda  - 6f(\phi ){H^2} - 6{H^2}\xi  - 2f(\phi )H' - 2\xi H' - 5H\Psi ' - {\Psi ^{\prime \prime }}\geq0.$
\end{itemize}
We will assume the vacuum solution e.g. $\rho=0$ and $p=0$ for investigation of these energy conditions.
\section{Different solutions}
In this section, we study evolution of all energy conditions for different solutions, power law solution as well as de Sitter solution.
\subsection{Power Law solutions}
Here, we assume power-law solutions with $H=J/t$ where $j$ is non zero constant and solving the following scalar field equation
\begin{equation}
\begin{split}
\phi''  &=  - 3H \phi'  - 6\left( { H' + 2{H^2}} \right){\mkern 1mu} ,\\
\xi''  &=  - 3H \xi'  - 6\left( { H' + 2{H^2}} \right){f_{,\phi }}(\phi ){\mkern 1mu} ,
\end{split}
\end{equation}
We supposing $f = {f_0}{e^{\alpha \phi }}$ and solving the above field equations, we get
\begin{equation}
\phi  = \frac{{{t^{1 - 3j}}C_1}}{{1 - 3j}} - \frac{{6j\left( {2j - 1} \right)\log (t)}}{{ - 1 + 3j}} + C_2,
\end{equation}

\begin{align}
\xi  &= \frac{1}{{((3j - 1)( - 1 - 6j(\alpha  - 1) + 3{j^2}(4\alpha  - 3)))}}\left[ {{t^{ - \frac{{3j\left( {1 + j\left( { - 6 + 4\alpha } \right) + 3{j^2}\left( {3 + 4\alpha } \right)} \right)}}{{{{\left( {1 - 3j} \right)}^2}}}}}} \right.\nonumber\\
&{(\frac{{\alpha {t^{1 - 3j}}C_1}}{{3j - 1}})^{\frac{{1 - 6j\left( {1 + \alpha } \right) + {j^2}\left( {9 + 42\alpha } \right)}}{{{{\left( { - 1 + 3j} \right)}^3}}}}}\left\{ {6{{\rm{e}}^{C_2\alpha }}} \right.{f_0}\alpha {(\frac{{\alpha {t^{1 - 3j}}C_1}}{{3j - 1}})^{ - \frac{{1 - 6j + 36{j^3}\alpha  + 3{j^2}\left( {3 + 4\alpha } \right)}}{{{{\left( { - 1 + 3j} \right)}^3}}}}}\nonumber\\
&j\left( {2j - 1} \right){t^{ - \frac{{6j\left( {1 + \alpha } \right)}}{{{{\left( {1 - 3j} \right)}^2}}}}}((3j - 1){t^{\frac{{3j\left( {3 + 9{j^2} + 2j\left( { - 3 + 7\alpha } \right)} \right)}}{{{{\left( {1 - 3j} \right)}^2}}}}} + C_1{t^{\frac{{1 + {j^2}\left( {9 + 42\alpha } \right)}}{{{{\left( {1 - 3j} \right)}^2}}}}}\alpha )\nonumber\\
&\Gamma (\frac{{6j\left( { - 1 + 2j} \right)\alpha }}{{{{\left( {1 - 3j} \right)}^2}}},\frac{{\alpha {t^{1 - 3j}}C_1}}{{3j - 1}}) + {t^{ - \frac{{6j\left( {1 + \alpha } \right)}}{{{{\left( {1 - 3j} \right)}^2}}}}}{(\frac{{\alpha {t^{1 - 3j}}C_1}}{{3j - 1}})^{ - \frac{{1 + 3j + 66{j^2}\alpha  + 9{j^3}\left( {3 + 8\alpha } \right)}}{{{{\left( { - 1 + 3j} \right)}^3}}}}}\nonumber\\
&\left\{ { - (C_3} \right.{{\rm{e}}^{\frac{{6j\left( { - 1 + 2j} \right)\alpha \left( {\left( { - 1 + 3j} \right)\log \left[ t \right] + \log \left[ {\frac{{\alpha {t^{1 - 3j}}C_1}}{{3j - 1}}} \right]} \right)}}{{{{\left( {1 - 3j} \right)}^2}}}}}{t^{\frac{{1 + {j^2}\left( {9 + 42\alpha } \right)}}{{{{\left( {1 - 3j} \right)}^2}}}}}{(\frac{{\alpha {t^{1 - 3j}}C_1}}{{3j - 1}})^{\frac{{9j\left( {1 + {j^2}\left( {3 + 4\alpha } \right) + j\left( { - 1 + 6\alpha } \right)} \right)}}{{{{\left( { - 1 + 3j} \right)}^3}}}}}\nonumber\\
& - C_4( - 1 + 3j){t^{\frac{{3j\left( {3 + 2\alpha  + j\left( { - 6 + 4\alpha } \right) + 3{j^2}\left( {3 + 4\alpha } \right)} \right)}}{{{{\left( {1 - 3j} \right)}^2}}}}}{(\frac{{\alpha {t^{1 - 3j}}C_1}}{{3j - 1}})^{\frac{{3j\left( {3 + 2\alpha  + j\left( { - 3 + 8\alpha } \right) + 3{j^2}\left( {3 + 8\alpha } \right)} \right)}}{{{{\left( { - 1 + 3j} \right)}^3}}}}})\nonumber\\
&( - 1 - 6j( - 1 + \alpha ) + 3{j^2}( - 3 + 4\alpha )) - 6C_1{{\rm{e}}^{C_2\alpha }}{f_0}j( - 1 + 2j){t^{\frac{{1 + {j^2}\left( {9 + 42\alpha } \right)}}{{{{\left( {1 - 3j} \right)}^2}}}}}\nonumber\\
&{\alpha ^2}{(\frac{{\alpha {t^{1 - 3j}}C_1}}{{3j - 1}})^{\frac{{1 + 36{j^3}\alpha  + 18{j^2}\left( {1 + 3\alpha } \right)}}{{{{\left( { - 1 + 3j} \right)}^3}}}}}\left. {\left. {\left. {\Gamma (\frac{{1 - 6j(1 + \alpha ) + 3{j^2}(3 + 4\alpha )}}{{{{(1 - 3j)}^2}}},\frac{{\alpha {t^{1 - 3j}}C_1}}{{3j - 1}})} \right\}} \right\}} \right].
\end{align}

Here $\Gamma(i,j)$ is incomplete gamma function.
In case of power law solution, we check the viability of the range different parameters. First, we find $\rho^{eff}>0$ and $\rho^{eff}+p^{eff}>0$ which constrained the parameters $t$, $C_1$ and $j$.\\

\begin{table*}[t]
\footnotesize
\centering \caption{Some of the constraint over parameters.}
\begin{tabular}{|c| c| c| c| }
\hline Label &  parameter 1 & parameter 2 & parameter 3  \\
\hline
\hline
&	$t$ & $j$ & $C_1>1$\\
$\rho^{eff}>0$&$t>1$ & $j=1$ & $C_1=\pm$ \\
& & $j>1$ & $C_1>0$\\
\hline
 &$t$ & $j$ & $C_1$\\
&$t>1$ & $j=1$ & $1 < C_1 < 6$\\
$\rho^{eff}+P^{eff}>0$& & $j=2$ & $1 < C_1 < 9$\\
& & $j=3$ & $1 < C_1 < 12$\\

\hline
& $t$&$j$&$C_1$\\
$\rho^{eff}-P^{eff}>0$& $t=1$ & & $0<C_1<3$\\
& $t=2$ &$j=1$ & $0<C_1<22$\\
& $t=3$ & & $C_1>0$\\
\hline
\end{tabular}
\label{tbl-2}
\end{table*}

we see that $\rho^{eff}>0$ impose the constraint over constant e.g.  $C_1>1$.
If we fix $j=1$ then $\rho^{eff}>0$ for both positive as well as negative values of $C_1$ but if we increase $j>1$ then $C_1$ should be any positive value. And by increasing $t$ then $\rho^{eff}$ becomes near to zero.\\

For the validity of $\rho^{eff}+p^{eff}$, the parameter $1<C_1<6$ with $t>1$ and $j=2$ while for $j=3$, one need $1<C_1<9$.\\
Similarly for $j=4$, the validity required the bound over parameter $1<C_1<12$ and so on.\\
$\rho^{eff}-p^{eff}$ for constant $j=1$, with $t=1$ required $0<C_1<3$ but for $t=2$ then $0<C_1<22$ and for $t>2$ then $C_1$ can take any positive values. Some of the possible values for the validity are:\\
i) For $j=2$ with $t=1$ then $0<C_1<8$,\\
ii) For $j=2$ with $t=2$ then $0<C_1<7$,\\
iii) For $j=2$ with $t=3$ then $0<C_1<9$,\\
iv) For $j=2$ with $t=4$ then $0<C_1<19$.\\
So on.\\

Furthermore, we check the behavior of $\rho^{eff}+3p^{eff}$ by fixing the $C_1$ to be a constant and vary $t$ along with $j$. we see that, for the validity of $\rho^{eff}+3p^{eff}$ imposed some constraint like:\\
i) For $t=1$ with $j=1$ required $C_1=0.5$.\\
ii) For $j=2$ with $t=2$ then $C_1=7.6$,\\
iii) For $j=2$ with $t=3$ then $C_1=15.43$,\\
iv) For $j=3$ with $t=4$ then $C_1=11.21$,\\
 And so on.\\
\begin{figure}\centering
\epsfig{file=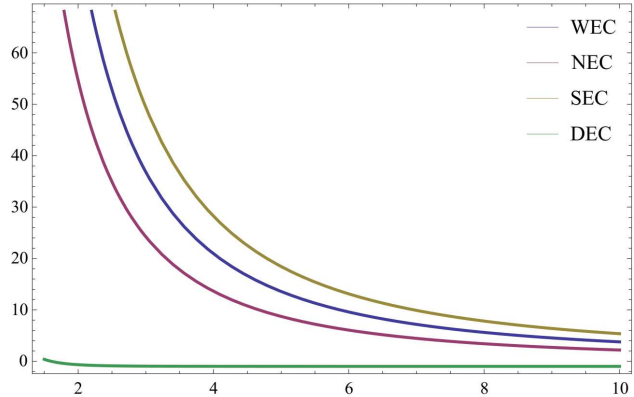,width=0.5\linewidth}
\caption{Validity of Energy conditions under Power law solutions.}\label{p11}
\end{figure}
The different Energy conditions are plotted against $t$ as shown in Fig. \ref{p11}.

\subsection{De Sitter with constant function solution}
In this section, we consider de Sitter solution along with constant function, as scalar field equations are
\begin{equation}
\begin{split}
\phi''  &=  - 3H \phi'  - 6\left( { H' + 2{H^2}} \right){\mkern 1mu} ,\\
\xi''  &=  - 3H \xi'  - 6\left( { H' + 2{H^2}} \right){f_{,\phi }}(\phi ){\mkern 1mu} ,
\end{split}
\end{equation}
Now assuming $H=H_0$ along with $f = f_0$, where $f_0$ is constant and solving the above field equation, we get the expression for both the scalar fields $\phi$ and $\xi$.
\begin{equation}
\phi  = C_2 - 4H_0t - \frac{{{{\rm{e}}^{ - 3H_0t}}C_1}}{{3H_0}}
\end{equation}
And
\begin{equation}
\xi  = C_4 - \frac{{{{\rm{e}}^{ - 3H_0t}}C_3}}{{3H_0}}
\end{equation}
Here we have four constant parameters $C_1$, $C_2$, $C_3$ and $C_4$.

$\rho^{eff}>0$ imposed the following constraint over these constant:
The product of $C_1$ and $C_3$ should be a negative constant, we assume $C_1$ to be a negative one along with negative value of $C_4$.
The validity of $\rho^{eff}+p^{eff}>0$ also required $C_1<0$ and $C_4<0$.

The dominant energy condition $\rho^{eff}-p^{eff}>0$ imposed the constraint over $C_4$, e.g. $C_4<-0.87068$. and the strong energy condition imposed the constraint as, $C_1<0$ or $C_3<0$ along with all positive value range of $C_4$ while there are also some regions where strong energy conditions are valid for $C_4<0$.
\begin{figure}\centering
\epsfig{file=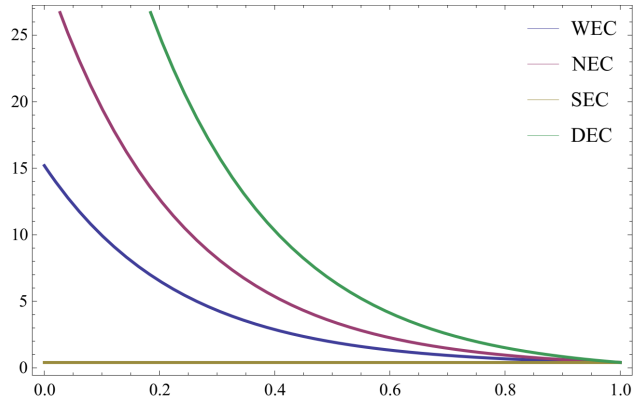,width=0.5\linewidth}
\caption{Validity of Energy Conditions under de Sitter solution having constant function.}\label{p22}
\end{figure}
The different Energy conditions are plotted against $t$ as shown in Fig. \ref{p22}.
In all these energy conditions, the constant $C_2$ has no role.
\subsection{De Sitter with linear function solution}
Furthermore, we assume de Sitter solution with a function $f =  \phi$. Now by solving field equations, we get the same expression for both the scalar fields $\phi$ and $\xi$.
\begin{equation}
\phi  = C_2 - \frac{{C_1{{\rm{e}}^{ - 3H_0t}}}}{{3H_0}} - 4H_0t
\end{equation}
And
\begin{equation}
\xi  = C_4 - \frac{{{{\rm{e}}^{ - 3H_0t}}C_3}}{{3H_0}} - 4H_0t
\end{equation}
In this case, we have four constant parameters $C_1$, $C_2$, $C_3$ and $C_4$.\\
For the validity of $\rho^{eff}>0$ :\\
i) the constant $C_1$ and $C_3$ should be opposite sign with the negative values of both $C_2$ and $C_4$.
ii) Weak energy condition is also valid for $C_2=-C_4$ along with $C_1.C_3=-C$, where $C$ is constant.
iii) Similarly, for small positive values of both $C_2$ and $C_4$ guaranty the Weak energy condition.\\

The validity of $\rho^{eff}+p^{eff}>0$ required $C_1<0$ or $C_3<0$ having large value for small $t$. e.g. $C_1.C_3=-308$ for $0<t=1$.

The dominant energy condition $\rho^{eff}-p^{eff}>0$ imposed the following possible constraint over parameters:\\
i)	For small positive values of $C_2$ and $C_4$,\\
ii)	For $C_2=-C_4$,\\
iiii) For any negative values of both $C_2$ and $C_4$.
While strong energy condition imposed the constraint over parameters as, for small $t$, the validity required $C_1<0$.\\

\begin{figure}\centering
\epsfig{file=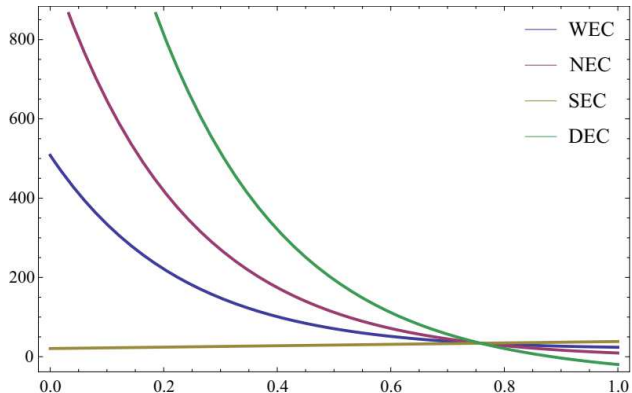,width=0.5\linewidth}
\caption{Validity of Energy Conditions under de Sitter solution having linear function.}\label{p33}
\end{figure}
The different Energy conditions are plotted against $t$ as shown in Fig. \ref{p33}.

\section{Summary}
In this paper we have studied a non-local gravity models given by the action \ref{action} which is constructed by adding the term $f(\Box^{-1}\mathcal{R})$ to the Einstein-Hilbert action. The $\Box^{-1}$ operator is the formal inverse of the Albertian in a scalar representation which can be expressed as a torsion with the retarded Green's function. The non-local distortion function $(\Box^{1}\mathcal{R})$ can be selected to reproduce the cosmology in the $\Lambda$CDM background. Ricci $\mathcal{R}$ disappears during the dominance of radiation, and $\Box^{-1}\mathcal{R}$ can not begin to grow until the beginning of the matter dominance while its growth becomes logarithmic by nature. Furthermore, the great advantage of this category of models is the release of the late-time acceleration by the transition from radiation domination.\\
Using this class of non-local gravity model with the localization procedure, we introduce two scalar fields and get the corresponding local gravity model. Furthermore, we have looked for two different types of solution, i) power-law and ii) de Sitter solutions. With the help of these solutions, we investigate the different type of energy conditions.
So far we have extracted the known Null energy conditions, the weak energy conditions, the strong as well as the dominant energy conditions of the non-local gravity model given by the action \ref{action}. In addition, in order to better understand the meaning of energy conditions, we have illustrated the evolution of the four energy conditions through different plots.\\

In case of power law solution we assumed $f=f_0e^{\alpha\phi}$ while for de Sitter solution we took $f=f_0$ and $f=\phi$, we see that all energy conditions are valid.
In general relativity, the power law solutions read $j=1/2$ for the dominant epoch of radiation and the solution with $j= 2/3$ for the matter dominant era. In this paper, we discuss the different values of $j$ where the energy conditions are valid.\\
The solutions obtained in this paper will be helpful for designing a realistic and physical model which is responsible for the accelerating nature of universe.Theoretical research may then lead to some qualitative results compared to the discussion of pure gravity. It will be implemented elsewhere.\\
It is noteworthy to investigate the existence of exact power solutions that correspond to different phases of early universe. These solutions are of particular importance because they are in a FLRW background that represents all the possible cosmic evolutions, such as the dark energy era, matter dominant or radiation dominant eras.\\
The non-local gravity has attracted considerable interest that depicts gravitational interactions, different from the more established theory of general relativity.

\medskip

%\begin{thebibliography}{72}

\end{document}